\documentclass[twocolumn,preprintnumbers,amsmath,amssymb]{revtex4}

\usepackage{graphicx}
\usepackage{dcolumn}
\usepackage{bm}

  \usepackage{latexsym}
  \usepackage{epsf}
  \usepackage{amssymb}

  \usepackage{amsmath}
  \usepackage{slashed,epsfig}

  \usepackage{bbm}
  \usepackage{bbm}
  \usepackage{amsmath,amssymb,amsthm}
  \usepackage{pst-all}

\begin{document}

\title{Racing through the swampland: \\de Sitter uplift vs weak gravity}

\author{Jakob Moritz$^1$,  Thomas Van Riet$^{2}$ }

\vspace{1.2cm}

\affiliation{{\small\slshape
$^1$ Deutches Electronen-Synchrotron, DESY, Notkestra$\beta$e 85,\\ 22607 Hamburg, Germany. \\
$^2$ Instituut voor Theoretische Fysica, K.U. Leuven,\\
Celestijnenlaan 200D, B-3001 Leuven, Belgium   } \\
{\upshape\ttfamily  jakob.moritz@desy.de\,\quad  thomas.vanriet@fys.kuleuven.be }
}

\begin{abstract}
We observe that racetrack models for moduli stabilization are in tension with strong forms of the Weak Gravity Conjecture (WGC). Moreover, recently, it was noted that controlled KKLT-type de Sitter vacua seem to require a racetrack fine-tuning of the type introduced by Kallosh and Linde. We combine these observations and conclude that the quests for realizing parametrically large axion decay constants and controlled de Sitter vacua are intimately related. Finally, we discuss possible approaches to curing the conflict between the racetrack scheme and the WGC.

\end{abstract}
\preprint{DESY 18-068}

\maketitle

	\section{Introduction}
	The cosmological constant problem remains as one of the most tantalizing problems in theoretical physics. Perhaps even more disturbing than the tiny observed value of the cosmological constant is its positive sign which is notoriously difficult to obtain in controlled string compactifications.
	
	Realizing de Sitter vacua in string theory is difficult, in part because all moduli need to be stabilized. This problem of moduli stabilization was solved long ago in the type IIB corner of string theory via the inclusion of three form fluxes to stabilize complex structure moduli \cite{Dasgupta:1999ss, Giddings:2001yu} and the leading non-perturbative corrections to stabilize also the volume modulus \cite{Kachru:2003aw, Balasubramanian:2005zx}. Recently this has been questioned \cite{Sethi:2017phn}, but we will work within the assumption that this way of stabilizing moduli is successful.
	
	However, given a controlled (i.e. supersymmetric) scheme of full moduli stabilization the possibility to `lift' these vacua to (non-supersymmetric) de Sitter vacua in a controlled way is not automatic. Using reasonable assumptions about the form of the potential of SUSY breaking sources (e.g. anti-branes or IASD fluxes), a successful lift of the SUSY AdS vacua of KKLT \cite{Kachru:2003aw} to de Sitter was claimed. While there are other classes of proposed stringy de Sitter vacua, the existence of the KKLT type de Sitter vacua can be considered an important argument for the existence of a vast landscape of de Sitter vacua in string theory.
	
	However, recently it was shown that the simplest examples of this type do not survive once the non-trivial interplay between the SUSY breaking source and the non-perturbative effects is taken into account \cite{Moritz:2017xto}. This was shown in a $10D$ framework and matched with subtle $\mathcal{O}(1)$ corrections to the antibrane potential in the language of $4D$ EFT. These corrections are only $\mathcal{O}(1)$ in magnitude but induce an exponential functional dependence of the antibrane potential on the volume modulus which is enough to prevent a successful uplift to de Sitter. It was also realized that models of supersymmetric volume stabilization that can accommodate supersymmetric Minkowski vacua with finite moduli masses are indifferent to these corrections and could hence generically allow for de Sitter uplifts.
	
	Given the remarkable robustness of such models on the one hand and the lack of explicit realizations on the other we find it interesting to ask whether such models can be constrained using general expectations about the properties of quantum gravity.

	\section{Constraints on uplifting}
	Uplifting AdS vacua to dS vacua is argued to be possible when an AdS vacuum  sits in the minimum of a sharply peaked potential and if there exist SUSY-breaking terms whose moduli-dependence is slowly varying compared with the sharply peaked AdS potential. If the two are combined, then the original AdS vacuum can be lifted without destroying the local minimum, on the premise that the magnitude of the SUSY-breaking term is tunable in size. The KKLT proposal \cite{Kachru:2003aw} suggests to use anti-D3 branes, which have an energy that depends polynomially on the K\"ahler modulus, $T$, via $V\sim Re(T)^{-2}$.   The KKLT AdS vacuum originates from exponential terms in the superpotential, $W = W_0 + A\exp(-aT)$\footnote{Here, $2\pi/a$ is the dual Coxeter number of a confining gauge group.}, and is indeed sharply peaked. The combined potentials are: 
	\begin{equation}\label{V}
	V(T) = e^{K/M_P^2}(g^{T\bar{T}}|D_T W|^2 - 3|W|^2/M_P^2) + \frac{\alpha^2} {12 Re(T)^{2}}\,.
	\end{equation}
	with $K$ the tree level K\"ahler potential,
	\begin{equation}\label{Kahler}
	K= -3M_P^2\ln(T+\bar{T})\, ,	
	\end{equation} 
	and $\alpha^2$ a positive number proportional to the warped anti-D3 tension. As usual we have assumed that the complex structure moduli have been integrated out in a non-SUSY GKP vacuum with finely-tuned small tree-level GVW superpotential $W_0$.
	 
	The tunability of the SUSY-breaking term $\alpha^2$ rests on the existence of highly warped throats inside compact Calabi-Yau spaces. Anti-D3 branes are dynamically attracted to the bottom of such throats, such that their tension is redshifted. It is well known that the amount of warping can be exponentially tuned \cite{Giddings:2001yu} such that the string scale tension can be brought to genuine low energy values. Despite the beautiful logic behind this reasoning, there are potential caveats. We focus on a particularly severe one which was only discovered recently \cite{Moritz:2017xto}. 
	
	In the uplift procedure it was assumed that the anti-D3 brane does not influence the non-perturbative corrections such that the uplift term can simply be added to the potential. The problem pointed out in \cite{Moritz:2017xto} is that this does not hold and that there is a non-trivial interplay between the effects that conspire against the construction of a dS vacuum. Reference \cite{Moritz:2017xto} has verified this directly in 10 dimensions.  For simplicity, we only state a possible interpretation of the problem in a 4-dimensional language (a variant of the one presented in \cite{Moritz:2017xto}) instead of presenting the $10D$ argument.
	
	In presence of non-perturbative volume stabilization, the anti-brane potential can receive non-perturbative corrections of the form
	\begin{equation}
	\delta V_{\overline{D3},np}\propto e^{-2a \text{Re}T}\, ,
	\end{equation}
	that can in principle compete with the (small) classical contribution\footnote{The uplift potential can alternatively be parametrized using the recent description of anti-brane SUSY breaking within $N=1$ supergravity \cite{Kallosh:2014wsa, Bergshoeff:2015jxa, Kallosh:2015nia} by making use of constrained superfields \cite{Volkov:1973ix, Komargodski:2009rz}.}.
	
	If the coefficient of such a correction is not suppressed by warping, and the coefficient is not numerically small, there exist no dS vacua that lie in a regime where $\alpha'$ corrections to the K\"ahler potential can be neglected: When the classical warp factor is tuned to obtain a dS vacuum, the $(mass)^2$ of the $T$-modulus is reduced by an entire volume factor. From a purely $4D$ point of view, such corrections may seem like a conspiracy but the $10D$ computation carried out in \cite{Moritz:2017xto} points precisely to it. There is no proof that this interpretation is correct and what follows in the rest of this paper does not rest on it.
	
	What counts in the following is the observation of \cite{Moritz:2017xto} that there seems to be a way to circumvent this problem. Whenever multiple non-perturbative effects are present, the $10D$ nogo theorem cannot be derived. The so called ``racetrack model" (see for instance \cite{Kaplunovsky:1997cy,Kallosh:2004yh}):
	\begin{equation}\label{racetrack}
	W = W_0 + A\exp(-a T) + B \exp(-bT)\,,
	\end{equation}
	is an example of this type. A possible source for two such exponential terms as the leading order non-perturbative effects are gaugino condensates on two stacks of D7 branes wrapping 4-cycles in the same homology class but at sufficiently large distances such that the gauge group induced by the 7-brane stacks factorizes. The reason that such models are not contained in the nogo theorem is the possible existence of a fine-tuning of the cosmological constant in the SUSY AdS vacuum prior to uplift. There is a tuning of the coefficients $A,B,a, b$ that brings the AdS vacuum energy to zero, while preserving finite masses for the $T$ modulus.
	In such a situation sufficiently small SUSY-breaking effects necessarily lead to dS vacua.
	
	In the rest of the paper we argue that such models can be challenged from two independent directions: 
	\begin{enumerate}
		\item Given the racetrack alignment of the dual Coxeter numbers of the gauge groups, $a\approx b$, for reasonable values of the parameters $W_0,A,B$, any reasonably strong form of the weak gravity conjecture (WGC) \cite{ArkaniHamed:2006dz} is violated, which some believe to be a fundamental principle of quantum gravity: A parametrically long axion direction emerges.
		\item Even taking for granted that all the gauge theory parameters assume the values required for ``racetracking", it is not obvious that eq. \eqref{racetrack} is even the correct \textit{form} of the IR superpotential \cite{Dine:1999dx}.
	\end{enumerate}
	We will outline these points in the following.
	\section{The WGC and axions}
	It is believed that some form of the \textit{weak gravity conjecture} (WGC) \cite{ArkaniHamed:2006dz} holds in all consistent effective field theories that arise as the low energy limit of a genuine string compactification. The (``electric") statement in its original form is that for a $U(1)$ gauge theory there should exist a particle of charge $q$ and mass $m$ that satisfies
	\begin{equation}
	q \gtrsim m\, ,
	\end{equation}
	in Planck units. If satisfied, extremal black holes can decay by emitting this particle. An early \textit{strong form} of this conjecture is that this particle has to be the lightest one in the charged spectrum. While strictly speaking counter-examples exist \cite{Heidenreich:2016aqi}, to the best our knowledge there does not exist a simple and controlled example that violates the statement parametrically \footnote{See however ref. \cite{Hebecker:2015rya}.}. 
	
	In string theory, the gauge theory statement can be extended via T-duality to $p$-form potentials coupled to $(p-1)$-branes. For axions ($0$-forms) coupled to instantons (($-1$)-branes), the statement is that there should exist an instanton with ``charge" $q/f$ and euclidean instanton action $S_E$ that satisfies
	\begin{equation}
	q/f \gtrsim S_E\, .
	\end{equation}
	Here, $f$ is the axion decay constant.
	
	We refer the reader to \cite{Rudelius:2014wla, Rudelius:2015xta, Bachlechner:2015qja, Brown:2015iha, Brown:2015lia, Junghans:2015hba} and references therein for a subset of the original literature on the extension of the WGC to axions and instantons.
	
	Since a controlled instanton expansion (i.e. in the ``dilute instanton gas" approximation) requires $S_E\gtrsim 1$, one concludes that the axion potential contains a harmonic,
	\begin{equation}
	V(\phi)\sim e^{-S_E}(1-\cos (q\phi/f))+...\, ,
	\end{equation}
	that varies on sub-Planckian distances in field space. If only the weakest form of the WGC holds, this harmonic could be induced by an instanton that gives negligible contributions to the potential \cite{Rudelius:2015xta,Brown:2015iha}, and natural inflation would not be very much constrained. However, if a sufficiently strong form holds, natural inflation is ruled out for the case of a single axion. As explained in ref. \cite{Brown:2015iha}, the natural generalization of the WGC to multiple $U(1)$'s \cite{Cheung:2014vva} also forbids axion decay constant enhancement using many-axions as in N-flation \cite{Dimopoulos:2005ac,Bachlechner:2015qja,Heidenreich:2016aqi} or alignment mechanisms \cite{Kim:2004rp,Palti:2015xra}.
	
	In the following we shall extrapolate the statement of the WGC to axions that obtain a periodic potential through \textit{any} non-perturbative effects (not necessarily instantons). We will work with the requirement that there should always exist a non-perturbative effect that induces an axion-potential that varies on sub-Planckian distances in field space. A strong form will be that this effect is non-negligible.
	
	\section{The WGC and de Sitter uplifts}
	
	The reason why we expect the WGC to constrain de Sitter uplifts is that the volume modulus is always accompanied by an axionic partner which in type IIB string theory can be thought of as the integral of the RR $4$-form over the $4$-cycle associated to the volume modulus. Given a stabilization mechanism for the volume we can hence inquire about the axion potential and whether or not it is consistent with the WGC. For instance, for the KKLT scenario with a single gaugino condensate $W=W_0 + A\exp(-aT)$, one can easily verify\footnote{The K\"ahler metric reads $g_{T\bar{T}}=\frac{3M_P^2}{(T+\bar{T})^2}$, so at fixed $\text{Re}(T)$ the canonically normalized axion is $\phi_c\sim M_P \text{Im}(T)/\text{Re}(T)$.} that
	\begin{equation}
	f/M_p \sim (a \text{Re}(T))^{-1} \,,
	\end{equation}
	up to order 1 coefficients. Hence, if higher order non-perturbative corrections can be ignored we necessarily have $a\text{Re}(T)>1$ such that $f$ is sub Planckian \footnote{Such corrections would likely be generated by higher-derivative corrections of the gauge theory \cite{Dine:1999dx}.}.
	
	We will focus on the racetrack stabilization \cite{Kallosh:2004yh}. The K\"ahler potential is given in (\ref{Kahler}) and the superpotential is given in (\ref{racetrack}) with 
		\begin{equation}
		a=\frac{2\pi}{N_1}\,,\quad b=\frac{2\pi}{N_2}\,.
		\end{equation}
	It is usually argued that this arises from gaugino condensation for the product gauge group $SU(N_1)\times SU(N_2)$ with gauge coupling set by the modulus $T$ (and no massless matter is assumed) \cite{Burgess:1998jh}. W.l.o.g in the following we take $b\geq a$, i.e. $N_1\geq N_2$.
	
	Splitting the real and imaginary (axionic) parts of $T=\sigma+ i \phi$, the  scalar potential $V(\sigma,\phi)$ reads \cite{BlancoPillado:2004ns}
	\begin{align}
	& V_0(\sigma)+V_1 (\sigma)\cos \left(\frac{2\pi}{N_1}\phi -\alpha\right)  +V_2(\sigma)\cos\left(\frac{2\pi}{N_2}\phi-\beta\right)\nonumber \\
	&+V_{1-2}(\sigma)\cos\left(\frac{2\pi(N_2-N_1)}{N_1N_2}\phi-\gamma\right)\, ,
	\end{align}
	with $\alpha\equiv \text{arg}(A\overline{W_0})$, $\beta\equiv \text{arg}(B\overline{W_0})$, $\gamma\equiv \text{arg}(A\overline{B})$ and coefficient functions
	\begin{align}
	\label{eq:harmonics}
	V_0(\sigma) =& \frac{1}{2\sigma^2}\Bigl(\frac{2\pi}{N_1}|A|^2(1+\frac{2\pi \sigma}{3N_1})e^{-\frac{4\pi}{N_1}\sigma} +\nonumber\\ 
	& \frac{2\pi}{N_2}|B|^2 (1 + \frac{2\pi\sigma}{3N_2}) e^{-\frac{4\pi}{N_2}\sigma} \Bigr)\,,\\
	V_1(\sigma)=& \frac{|AW_0|}{2\sigma^2}\frac{2\pi}{N_1}\, e^{-\frac{2\pi}{N_1}\sigma}\, ,\\
	V_2(\sigma)=& \frac{|BW_0|}{2\sigma^2}\frac{2\pi}{N_2}\, e^{-\frac{2\pi}{N_2}\sigma}\, ,\\
	V_{1-2}(\sigma)=&\frac{|AB|}{2\sigma^2}\Bigl(\frac{2\pi}{N_1}+\frac{2\pi}{N_2}+\frac{8\pi^2\sigma}{3N_2N_1}\Bigr)e^{-(\frac{2\pi}{N_1}+\frac{2\pi}{N_2})\sigma}\, .
	\end{align}
	There exist three distinct harmonics for the axion $\phi$ with coefficient functions $V_1,V_2$ and $V_{1-2}$. The last one has periodicity $\frac{N_1 N_2}{N_1-N_2}$ and if dominant leads to an effective axion decay constant
	\begin{equation}
	f_{\phi}/M_P\sim \frac{N_1 N_2}{N_1-N_2}\cdot \frac{1}{\sigma}\, .
	\end{equation}
	This is super-Planckian when $N\equiv N_1\approx N_2$ and $N<\sigma < N^2$, a regime where the fractional instanton expansion would naively seem to be under control. In order for this harmonic to be sub-dominant all the way down to the breakdown of the fractional instanton expansion at $\sigma \sim a^{-1}$ (i.e. $V_{1-2}\overset{!}{\leq} \text{max}(V_1,V_2)$) we would have to demand that
	\begin{equation}
	\label{conclusion1}
	|W_0|\gtrsim \text{min}(|A|,|B|)\, .
	\end{equation}
	In this case, the strong form of the WGC would hold all the way down to $\sigma\sim a^{-1}$. However, we would find such a strict bound on the flux number $W_0$ very surprising, in particular because it seems that no such bound can be derived for the single gauge group KKLT model. Moreover, explicit studies of the classical flux superpotential indicate that the genericity arguments for a small $W_0$ are valid \cite{MartinezPedrera:2012rs}.
	
	The virtue of this model would be that when the parameters $A, B, W_0$ of the model are tuned to satisfy
		\begin{equation}
		\label{tuning}
		-W_0=A\left(-\frac{N_2 A}{N_1 B}\right)^{N_2/(N_1-N_2)}+B\left(-\frac{N_2A}{N_1B}\right)^{N_1/(N_1-N_2)}\, ,
		\end{equation}
	there exists a SUSY Minkowski minimum at
		\begin{equation}
		\label{Tmin}
		2\pi T_0\equiv 2\pi (\sigma_0+i \phi_0) =\frac{N_1 N_2}{(N_1-N_2)}\log \left(-\frac{B N_1}{A N_2 }\right)\, ,
		\end{equation}
	and the mass of the volume modulus is finite (a corresponding minimum exists also when the relation \eqref{tuning} is detuned, but the vacuum will be of Anti-de-Sitter type.). In order for this minimum to lie at positive volume it is required that $|Bb|>|Aa|$. 
	
	If the tuning of eq. \eqref{tuning} holds, one has that $|W_0|\leq |Ae^{-a\sigma_0}|+|Be^{-b\sigma_0}|$. It then follows that 
	the WGC-type bound \eqref{conclusion1} cannot possibly be satisfied unless $e^{-a\sigma_0}\gtrsim \mathcal{O}(1)$. In other words, the Minkowski racetrack minimum would lie outside the (naive) validity of the controlled fractional instanton expansion\footnote{Note, that we assume there is no hierarchy between the coefficients $A,B$, i.e. $|A|/|B|=\mathcal{O}(1)$.}. As a consequence, if the WGC holds, the racetrack minimum cannot be used as a controlled starting point for uplifting to de Sitter space, i.e. there is no parametricallly controlled de Sitter uplift within the racetrack scheme.
	
	In the following section we will argue for ways to cure the conflict with the WGC in a different (from eq. \eqref{conclusion1}), and perhaps more drastic way.
	
\section{Discussion}
We have explained how racetrack schemes of moduli stabilization violate strong forms of the WGC. Let us first comment on the implications \textit{assuming} a strong form of the WGC must always hold in string theory: Together with the results of ref. \cite{Moritz:2017xto}, one may arrive at the suspicion that the KKLT construction does not result in de Sitter vacua even when the scheme of non-perturbative volume stabilization is extended beyond a single non-perturbative effect. In particular, adding anti-D3 branes to the SUSY KKLT vacua would lead to meta-stable non-SUSY AdS vacua at best. More speculatively, string theory might not harbor parametrically controlled de Sitter vacua in its weakly coupled corners \cite{Brennan:2017rbf,Danielsson:2018ztv}.

We now present an alternative line of argument that could lead to compliance with the WGC, without having to impose eq. \eqref{conclusion1}. To this end, let us critically review the derivation of \eqref{racetrack}. It is assumed that in the IR we are left with an $\mathcal{N}=1$ SUSY $SU(N_1)\times SU(N_2)$ pure Yang-Mills theory with (UV-)gauge coupling set by the volume modulus $T$. The Lagrangian is
\begin{equation}
\mathcal{L}=\frac{1}{32\pi}\text{Im} \int d^2\theta \, \,\sum_{a=1}^{2} \tau_a  \Tr{W_a^{\alpha}W_{a\,\alpha}}\, +\int d^4 \theta\,  K(T+\overline{T})+...\, ,
\end{equation}
with holomorphic gauge couplings $\tau_a=-i T$, gaugino superfield $W_{a,\alpha}=-i \lambda_{a,\alpha}+\mathcal{O}(\theta)$ and K\"ahler potential $K(T,\overline{T})\equiv K(T+\overline{T})$ (and we have omitted higher-derivative corrections).

Classically, this theory enjoys a $U(1)_R$ symmetry that acts as
\begin{equation}
\lambda_{1,2}\longrightarrow e^{i\alpha} \lambda_{1,2}\,,\quad
\chi_T \longrightarrow e^{-i\alpha} \chi_T\, ,
\end{equation}
where $\chi_T$ is the superpartner of the K\"ahler modulus $T$.

This $U(1)_R$ is anomalous and the action transforms according to
\begin{equation}\label{anom.trafo1}
\tau_a\longrightarrow \tau_a-N_{a}\frac{\alpha}{\pi}\, .
\end{equation}
This is a symmetry only if
\begin{equation}
\alpha=\frac{n}{g} \pi\, ,\quad n=0,...,2g-1 \, ,
\end{equation}
where $g\equiv \gcd(N_1,N_2)$ is the greatest common divisor of $N_1$ and $N_2$. The $U(1)_R$ symmetry is thus broken to its discrete subgroup $\mathbb{Z}_{2g}$ by gauge instantons. Clearly, unless $N_1=N_2$ it is not possible to cancel the anomaly through a shift in $T$ as is possible for a single gauge group or two (or more) gauge groups with identical dual Coxeter numbers (i.e. equal ranks for $SU(N)$ groups). 

As a result, it does not seem to be possible to uniquely determine the IR superpotential using holomorphy and demanding the correct $R$-charge of the superpotential. As has been pointed out by \cite{Dine:1999dx} arguments for the superpotential being the sum of two exponential terms treat $T$ as a background field.

In terms of the renormalization group (RG) invariant IR scales $\Lambda_a^{3N_a}\equiv \mu^{3N_1}\exp(2\pi i\tau_a(\mu))$, the transformation \eqref{anom.trafo1} amounts to assigning $R$-charge $2N_a$ to the IR scales. Specifying to $N_1=N_2+1$, the most general IR-superpotential that has the correct $R$-charge can be written as
\begin{equation}
W(\lambda,u)=\lambda \cdot c(u)\, ,
\end{equation}
where $\lambda\equiv \frac{\Lambda_1^{3N_1}}{\Lambda_2^{3N_2}},$ $u\equiv \frac{(\Lambda_2^{3N_2})^{N_1}}{(\Lambda_1^{3N_1})^{N_2}}$, and $c(u)$ is an arbitrary (multivalued) holomorphic function.

By taking the limits $\Lambda_a^{3N_a}\longrightarrow 0$, and matching with the single gauge group results, one concludes
\begin{equation}\label{limits}
c(u\longrightarrow 0)=N_1u^{\frac{1}{N_1}}\, ,\quad c(u\longrightarrow \infty)=N_2u^{\frac{1}{N_2}}\, .
\end{equation}
These are also the limits in which larger discrete subgoups $\mathbb{Z}_{2N_a}\subset U(1)_R$ are restored. The racetrack superpotential of eq. \eqref{racetrack} corresponds to
\begin{equation}
c_{rt}(u)=N_1u^{\frac{1}{N_1}}+N_2u^{\frac{1}{N_2}}\, ,
\end{equation}
and is by far not the unique function that satisfies \eqref{limits}. 

In the language of the variable $u$, the enhancement of the axion field range of the modulus $T$ that occurs with the racetrack proposal corresponds to the function $c(u)$ being single-valued holomorphic only in the variable $u^{\frac{1}{N_1N_2}}$. Determining the function $c(u)$ (or at least its monodromy around the origin), would be an important step toward deciding the viability of racetrack models and KKLT de Sitter uplift mechanisms. To us there are the following realistic possibilities:
\begin{enumerate}
	\item The IR superpotential of $SU(N_1)\times SU(N_2)$ in the presence of a perturbatively massless dilaton (K\"ahler modulus) does not reduce to the racetrack form of eq. \eqref{racetrack}. Rather, the function $c$ is holomorphic\footnote{when restricted to a sufficiently small patch around $u=0$.} in $u^{\frac{1}{\mathcal{O}(N)}}$. In this case, all phenomenological models based on it would have to be revisited\footnote{A subset of those being \cite{deCarlos:1992kox,BlancoPillado:2004ns,Lalak:2005hr,BlancoPillado:2006he}.}.
	\item The field theory superpotential is as in eq. \eqref{racetrack}, but there are gravitational corrections (such as gravitational instantons) that are non-negligible. Implications for racetrack model building could be very much model dependent.
	\item The IR superpotential is essentially given by the racetrack proposal, but small $W_0$ is not attainable when the gauge sectors are arranged as required (see eq. \eqref{conclusion1}).
	\item Only the weakest forms of the weak gravity conjecture hold.
\end{enumerate}
Whichever holds, we observe an unexpected relation between the ability to realize parametrically large axion decay constants and parametrically controlled de Sitter vacua. Success or failure to achieve the former, possibly determines the viability of the latter.
 
We leave a more detailed analysis of the IR superpotential of the racetrack scheme for future work.

\section*{Acknowledgments}
We thank Riccardo Argurio, Thomas Bourton, Ulf Danielsson, Mafalda Dias, Michael Dine, Jonathan Frazer, Shamit Kachru, Miguel Montero, Elli Pomoni, Ander Retolaza, Andreas Ringwald and Alexander Westphal for useful discussions. JM would like to thank the Instituut voor Theoretische Fysica at KU Leuven for the warm hospitality while this work was initiated. The work of TVR is supported by the FWO odysseus grant G.0.E52.14N and the C16/16/005 grant of the KULeuven. The work of JM is supported by the ERC Consolidator Grant STRINGFLATION under the HORIZON 2020 grant agreement no. 647995.

\bibliography{racetrackWGC}

\end{document}